# Strain effects on borophene: ideal strength, negative Possion's ratio and phonon instability

Haifeng Wang[1,2*], Qingfang Li[3*], Yan Gao[2], F. Miao[1], Xiang-Feng Zhou[4] and X. G. Wan[1]

[1]National Laboratory of Solid State Microstructures, School of Physics, Collaborative Innovation Center of Advanced Microstructures, Nanjing University, Nanjing 210093, China

[2]Department of Physics, College of Science, Shihezi University, Xinjiang 832003, China

[3]Department of Physics, Nanjing University of Information Science & Technology, Nanjing 210044, China

[4]School of Physics and Key Laboratory of Weak-Light Nonlinear Photonics, Nankai University, Tianjin 300071, China

Very recently, two-dimensional (2D) boron sheets (borophene) with rectangular structure has been grown successfully on single crystal Ag(111) substrates [Science **350**, 1513 (2015)]. The fabricated boroprene is predicted to have unusual mechanical properties. We performed first-principle calculations to investigate the mechanical properties of the monolayer borophene, including ideal tensile strength and critical strain. It was found that monolayer borophene can withstand stress up to 20.26 N/m and 12.98 N/m in **a** and **b** directions, respectively. However, its critical strain was found to be small. In **a** direction, the critical value is only 8%, which, to the best of our knowledge, is the lowest among all studied 2D materials. Our numerical results show that the tensile strain applied in **b** direction enhances the bucking height of borophene resulting in an out-of-plane negative Poisson's ratio, which makes the boron sheet show superior mechanical flexibility along **b** direction. The failure mechanism and phonon instability of monolayer borophene were also explored.

--------------------------------------------------------------------------------

* Correspondence and requests for materials should be addressed to Haifeng Wang (e-mail: whfeng@shzu.edu.cn); or to Qingfang Li (e-mail: qingfangli@nuist.edu.cn)

Boron is a fascinating element because of its chemical and structural complexity.

Although it is carbon's neighbor in the periodic table with similar valence orbitals, the electron deficiency prevents it from forming graphene-like planar structures. In spite of numerous theoretical proposals,[1-6] borophene had not been synthesized successfully till very recently on singly crystal Ag(111) substrates under ultrahigh-vacuun conditions.[7] The monolayer borophene with rectangular structure has shown some extraordinary properties,[2,7-9] including the anisotropic metallic character and unique mechanical properties. For example, it exhibits an extremely large Young's modulus of 398 GPa·nm along **a** direction,[7] which exceeds the value of graphene. The borophene shows great potential for applications in nano-scale electronic devices and micro-electro-mechanic systems (MEMS) due to these novel properties. Adventitious strain is almost unavoidable experimentally, therefore, it is highly desirable to explore the mechanical properties of borophene.

For 2D materials, the ideal tensile strength[10,11] is a crucial mechanical parameter which fundamentally characterizes the nature of the chemical bonding and elastic limit of the single- or few-layer thin films. So far, the elastic limit of many 2D materials, such as graphene,[12-14] $h$-BN,[15-19] $MoS_2$,[20-24] black phosphorene (BP)[25-28] and silicene,[29-33] have been characterized by the ideal tensile stress and critical strain. Compared to these materials, monolayer borophene is a stiffer material because of higher Young's Modulus.[7] In this work, we presented systematic analysis on the strain-induced mechanical properties of monolayer borophene, including the ultimate stress and critical strain, the change of bucking height, and the failure mechanism when approaching the limit strain, and compared them with other representative 2D materials: graphene, silicene, BP and MoS$_2$. We found that unlike graphene and silicene, which own hexagonal structure and show slight anisotropy of mechanical properties, monolayer borophene posses rectangular structure which brings about significantly anisotropic ultimate strengths and critical strains. The strong $\sigma$ bonds lying along **a** direction plays an important role in the mechanical properties of borophene, which not only results in the stiffness even rivals graphene[7] but also lead to the ideal tensile stress along **a** direction much larger than that of MoS$_2$,[20-22] BP[25] and silicene.[29-33] The critical strain of monolayer borophene is small, which is only 8% in **a** direction, to the best of our knowledge, this critical strain is the lowest among all studied 2D materials. So, borophene can be seen as a hard and brittle material. For the tension applied in **b** direction, the boron sheet shows superior mechanical flexibility and out-of-plane negative Poisson's ratio resulted from the bucking height of borophene increasing with strain, just like BP.[34] The negative Poisson's ratio of BP originates from its puckered structure, while for borophene, we found its negative Poisson's ratio mainly results from the weakening of the interlayer B$_1$-B$_2$ bonding with increasing **b**-axis strain. Furthermore, the failure mechanisms of borophene upon tension are found very similar with MoS$_2$,[20,22] in one direction it is attributed to elastic instability, while in the other direction the failure mechanism is phonon instability and such an instability is dictated by out-of-plane acoustical (ZA) mode.

Our first principle calculations were carried out with the Vienna *ab-initio* Simulation Package (VASP)[35] based on density functional theory (DFT). The Perdew-Burke-Ernzerh of (PBE) exchange-correlation functional[36] along with the projector-augmented wave (PAW) potentials was employed for the self-consistent total energy calculations and geometry optimization. The valence electronic configurations for rhomb were chosen as $2s^2p^1$. The kinetic energy cutoff for the plane wave basis set was chosen to be 500 eV. The Brillioun zone was sampled using a $25 \times 15 \times 1$ Monkhorst-Pack k-point grid. Atomic positions were relaxed until the energy differences were converged within $10^{-6}$ eV and the maximum Hellmann-Feynman force on any atom was below 0.001 eV/Å. A vacuum of 20 Å along **c** direction was included to safely avoid the interaction between the periodically repeated structures. The phonon spectrum was calculated using the PHONOPY code.[37] A $7 \times 5 \times 1$ supercell with $7 \times 5 \times 1$ k-mesh was used to ensure the convergence.

The theoretical stress-strain relation were predicted by following a standard method.[38,39] To compute the stress-strain relationship of **a**-direction, as defined in Fig. 1, we applied a series of incremental tensile strains on the rectangular unit cell along **a**-direction and relaxed the lattice along **b**-directions until the corresponding conjugate stress components was less than 0.01 GPa. The **b**-direction uniaxial stress state was solved analogously. For biaxial stress-strain calculations, equibiaxial tension was applied and then the B atoms in unite cell were fully relaxed. The engineering tensile strain is defined as $\varepsilon = (L - L_0)/L_0$, where $L$ is the strained lattice constants and $L_0$ is the original lattice constants, respectively. Currently the distance of interlayer in borophene structure can't be determined experimentally, we thus used the in-plane stress $f$ (2D force per length with unit of N/m) to represent the strength of the structure.[40] The 2D stress can be expressed by multiplying the Cauchy stresses (one of the outputs from VASP) and the thickness of the unit cell.

The optimized structure of monolayer borphene is shown in Fig. 1. The calculated lattice parameters of borophene are a=1.614 Å and b=2.866 Å, which is in good agreement with the experimental and theoretical results.[7,8] Borophene has a highly anisotropic crystal structure, and there is the bucking along the **b** direction with height $h = 0.911$ Å, while no corrugations along **a** direction are observed.

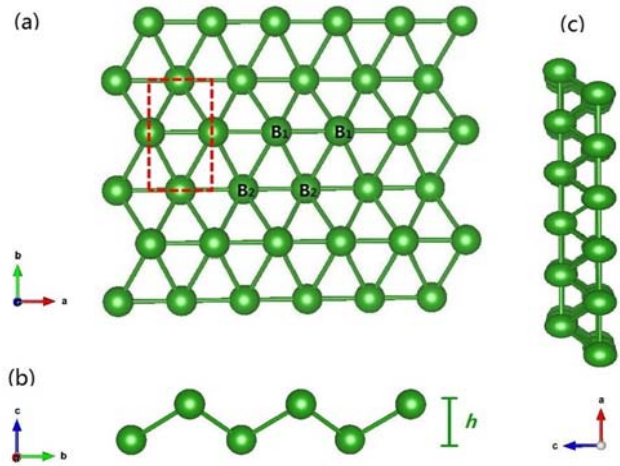

FIG. 1. (a) Top, (b and c) side views of the atomic structure of borophene. The two-atom orthorhombic unit cell framed by red dashed line in panel (a) is used to perform the stress-strain relationship calculations under uniaxial and biaxial tensions. $B_1$ and $B_2$ represent the two nonequivalent boron atoms in the unit cell.

Starting with the relaxed borophene structure, tensile strain is applied in either uniaxial or biaxial direction to explore its ideal tensile strength. Our calculated strain-stress relations of monolayer borophene are presented in Fig. 2. By fitting the initial stress strain curves based on linear regression up to 2% along **a** and **b** directions, we obtained the corresponding Young's modulus $E_a = 389$ N/m and $E_b = 166$ N/m, in excellent agreement with the results of previous literature.[7]

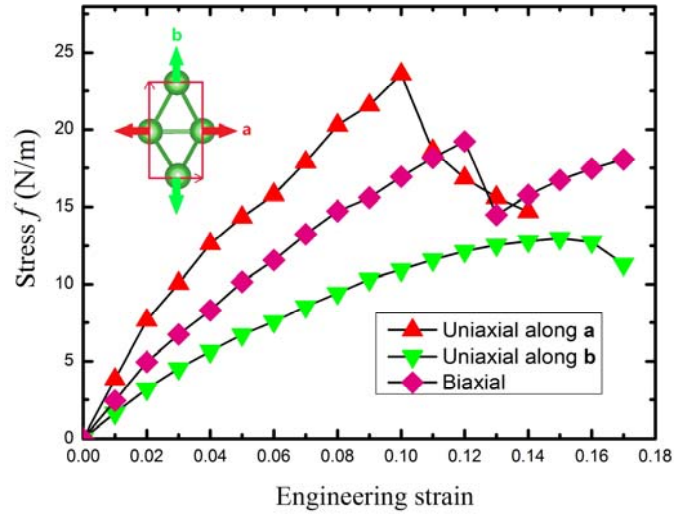

FIG. 2. Calculated stress-strain relationships of borophene under uniaxial **a**, **b** and biaxial tensions.

As shown in Fig. 2, the calculated strain-stress behaviors become nonlinear as the applied strain increases. The maximum stress for uniaxial tension in **a** direction is 24.0 N/m, and the corresponding critical strain is 0.10. The tensile strength value is much larger than that of $MoS_2$,[20-22] $BP$[25] and silicene.[29-33] It's not surprising because a strong $\sigma$ bonds is found lying along **a** direction.[2] For **b** direction tension, monolayer borophene demonstrates a smaller tensile strength of 12.98 N/m since only the slightly weaker multicenter bonds are involved.[2] On the other hand, borophene shows more superior flexibility when tension is applied along **b** direction with the critical strain of 0.15. For biaxial tension case, the curve has a maximum value of 19.21 N/m when the strain is applied to 0.12. Interestingly, the curve has a minor value at the strain of 0.13.

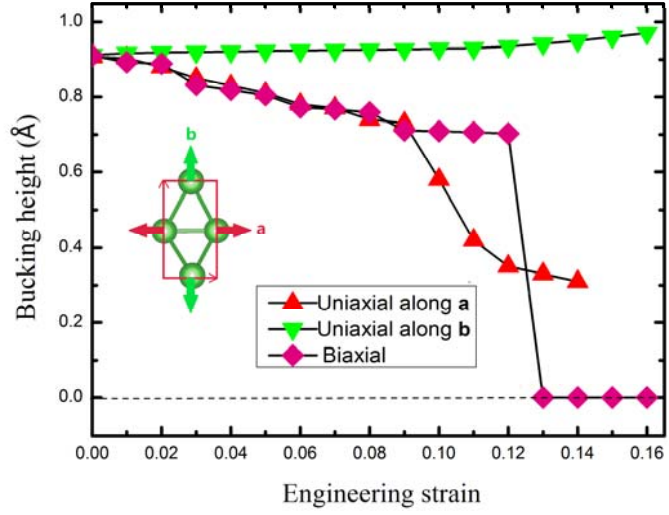

FIG. 3. The calculated dependence of buckling height of borophene on the three types of tensions.

The buckling height $h$ is an important parameter to characterize the corrugation of 2D materials.[25,26,29,32] The dependences of buckling height on the three types of tensions are shown in Fig. 3. It can be seen that the tension dependent buckling heights are anisotropic and non-monotonic. When tension is applied along **a** direction, the bucking height decreases significantly with increasing strain. This trend effectively flattens the pucker of single-layer 2D sheets, which significantly reduces the required strain energy. The buckling height under the biaxial tension decreases monotonously with increasing strain before the strain approaches the critical value (0.12), but it quickly decreases to zero as the strain continuously increases. That is to say, borophene turns to a graphene-like plane structure instead of the original bucking structure in this case. It's just the reason why the curve of stress-strain relationship under biaxial tension (Fig. 2) has a minor value corresponding to the strain of 0.13. But in fact, such a structural distortion can't happen since the phonon instability happens at less strain as will be discussed later.

More surprisingly, the buckling height under uniaxial **b** direction tension increases monotonically with increasing strain. It means that just like BP,[34] monolayer borophene also has an out-of-plane negative Poisson's ratio. J. Mannix *et al.* reported the in-plane negative Poisson's ration of borophene both along **a** and **b** direction (equal to –0.04 along **a** and –0.02 along **b**) within the strain range between -2% and 2%. Now we find when larger tensile strain was applied along **b** direction, monolayer borophene also has an out-of-plane negative Poisson's ratio. According to ref. [34], the negative Poisson's ratio of single-layer BP originates from its puckered structure, while for borophene, we think the out-of-plane negative Poisson's ratio mainly results from the weakening of the interlayer $B_1$-$B_2$ bonding with increasing **b**-axis strain, as can be seen from the valence electron density in Fig. 4 (a). Such phenomenon still comes from the strong $\sigma$ bonds of **a**-direction. Specifically, when tension is applied on **b**-direction, borophene is stretched in this direction, to accommodate the elongation in **b**-direction, borophene should contract in **a**-direction, but the $\sigma$ bonds lying on **a**-direction is so strong that enough contraction won't happen timely and effectively. Then the interlayer $B_1$-$B_2$ bond length becomes more and more large, and reasonably, the strength of $B_1$-$B_2$ bonding turns more and more weak. In comparison, as shown in Fig. 4 (b), the $B_1$-$B_2$ bonds shows normal contractive behavior and the $B_1$-$B_2$ bonding becomes more and more strong when tension was applied along **a** direction, so the bucking height of borophene will reasonably decrease in this case. It's not surprising since along **b**-direction of monolayer borophene, there is only weak multicenter bonds.

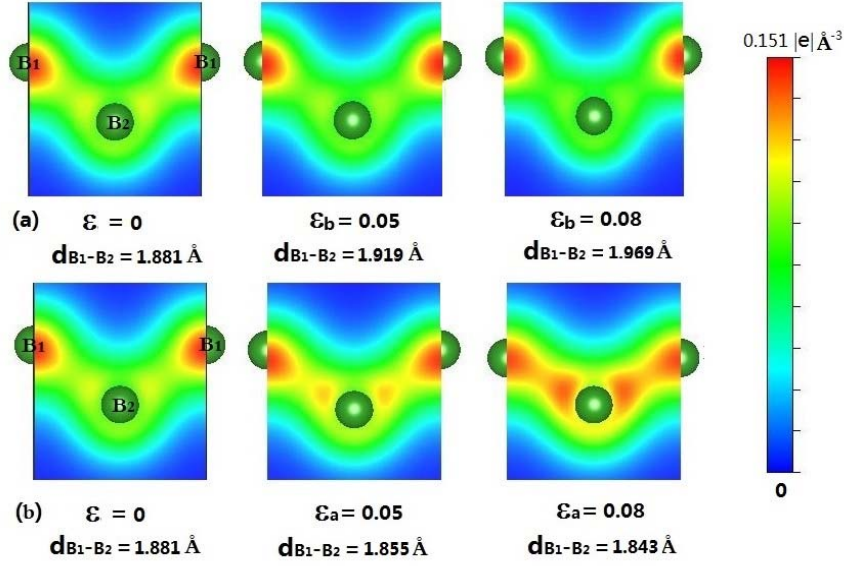

FIG. 4. Valence electron density of monolayer borophene along **a**-direction under different uniaxial (a) **b**-direction strains and (b) **a**-strains, and the corresponding changes of interlayer $B_1$-$B_2$ bond lengths.

The anisotropic behavior of the buckling height along **a** and **b** directions (Fig. 3) gives an interpretation for the discrepancy of flexibility of monolayer borophene. As been revealed by J. Kunstmann *et al*,[2] any flattening of the boron sheet would cause $p_y$ orbitals to interfere with the strong $\sigma$ bonds along **a** direction and eventually destroy them. From Fig. 3 we find **a** direction tension will significantly decrease the bucking height and flatten the borophene structure rapidly, thus the strong $\sigma$ bonds will be greatly interfered and corresponding the structure becomes instable. On the contrary, when the tension is applied along **b** direction, the bucking height will increase, thus the pure stong $\sigma$ bonds along **a** direction will be hold even at large tension. Therefore, the 2D boron sheet shows superior mechanical flexibility along **b** direction.

In fact, Fig. 2 only provides a rough indication of the strength of borophene upon different tension and it is important to verify whether the monolayer borophene remains stable before approaching the maximum stress. So, we carried out the phonon dispersion calculation in order to examine the stability conditions. The stress-free phonon spectrum is illustrated in Fig. 5 (a). The primitive cell of borophene contains two atoms, so there are three acoustic and three optical phonon branches. The longitudinal acoustic (LA) and transverse acoustic (TA) branches correspond to vibration within the plane, and the other one (ZA) corresponds to vibration out of plane. It should be noticed that the ZA branch of borophene has a small imaginary frequency along $\Gamma$-X direction, which indicates that the lattice may exhibit instability against long-wavelength transversal waves.[7,41,42]

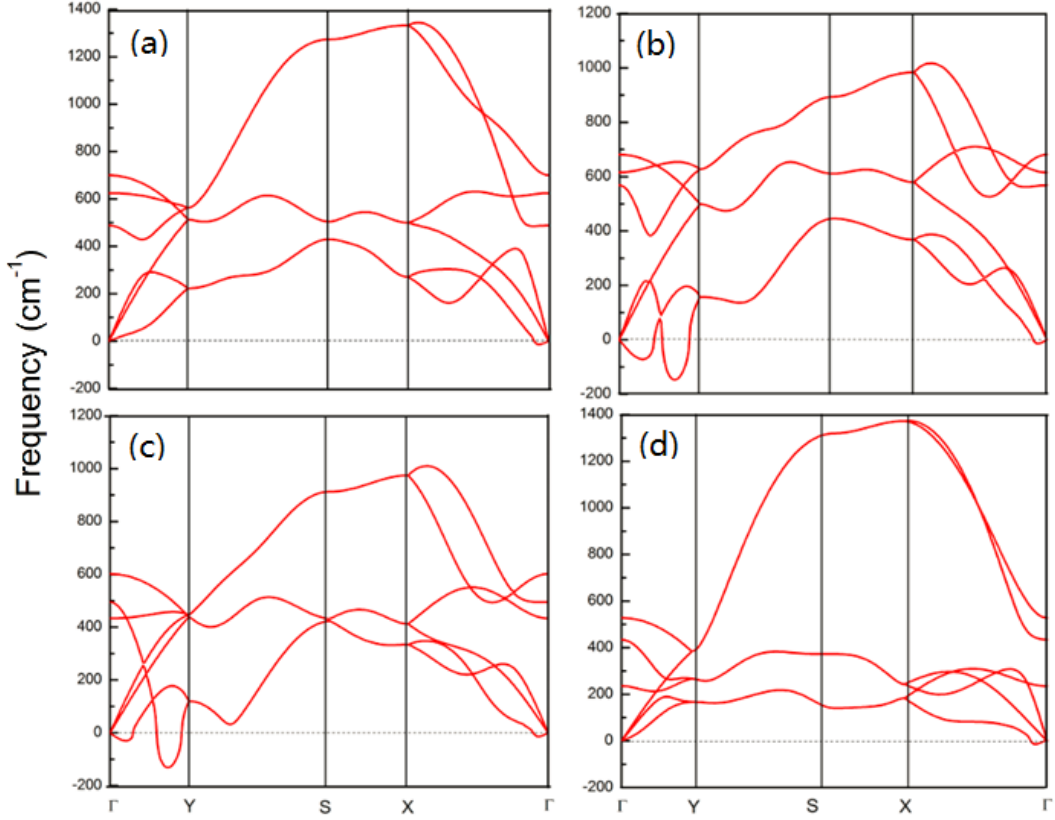

FIG. 5. Phonon dispersion at (a) stress-free state, (b) uniaxial tensile strain $\varepsilon_a = 0.08$, (c) biaxial tensile strain $\varepsilon_{biaxial} = 0.08$ and (d) uniaxial tensile strain $\varepsilon_b = 0.15$.

For the uniaxial tensile strain $\varepsilon_a = 0.08$ and biaxial tensile strain $\varepsilon_{biaxial} = 0.08$, a phonon branch has the negative (imaginary) frequencies along Γ-Y direction as shown in Fig. 5 (b) and (c). Examination of the eigenvectors of the unstable phonon modes shows that the soft mode is ZA branch, corresponding to vibration out of the plane. Interestingly, for both the uniaxial tension along **a** and biaxial tension, the single-layer borophene becomes dynamically unstable when the $B_1$-$B_1$ (and $B_2$-$B_2$) bong lengths (Fig. 1) are extended 8%, suggesting a lower limit for which the single-layer borophene remains dynamically stable along the out-of-plane direction. In contrast, such phonon mode remains stable under uniaxial tension along **b** direction when reaching the maximum tensile stress [Fig. 5 (d)]. That is to say, in this case, the failure mechanisms of borophene under uniaxial tensions are elastic instability.

Since the instability conditions and failure mechanism of borophene under three types of tension are determined, the ideal tensile strengths and critical strains can be confirmed. And it is interesting to compare the mechanical responses of single-layer borophene under tension with those of some other studied 2D materials. Table I summarizes the calculated ideal tensile strengths and critical strains of borophene along three loading directions. And we compared the critical values and mechanisms of borophene with that of graphene, silicene, BP and $MoS_2$ in Table I. It should be noticed that in ref. [12], ref. [22] and ref. [25], the ideal stresses are given in Chancy stress with unit of GPa, in order to facilitate the comparison with each other, here we unified these values as in-plane stress in unit of N/m. As shown in Table.1, graphene and silicene show slight anisotropy of ultimate strengths and critical strains because of their hexagonal structure, while for borophene and BP, the anisotropic crystal structures are responsible for their significantly anisotropic mechanical properties. The ideal tensile strengths of monolayer borophene are obviously larger than that silicene of and BP, while much smaller than that of graphene, due to the fact that the B-B bonds are stronger than Si-Si bonds and P-P bonds, while weaker than C-C bond in graphene. The corresponding strain along **a** direction (8%) is quite small, to the best of our knowledge, this critical strain of monolayer borophene may be the lowest among all of the studied 2D materials. For the tension applied in **b** direction, the boron sheet shows superior mechanical flexibility and out-of-plane negative Poisson's ratio resulted from the bucking height of borophene increasing with strain, just like

BP.[34] The failure mechanisms of graphene and silicene are similar, that is, under uniaxial and biaxial tensions their failure mechanisms are elastic instability. While the failure mechanisms of borophene upon tensions are found very similar with $MoS_2$,[20,22] in one direction it is attributed to elastic instability, while in the other direction the failure mechanism is phonon instability and such an instability is dictated by out-of-plane acoustical (ZA) mode.

TABLE I. Summary of the ideal tensile strengths, critical strains, and failure mechanisms of monolayer borophene under three strain paths and comparison with graphene, silicene and BP.

|  | Direction | $f$ (N/m) | $\varepsilon_c$ | Failure mechanism |
|---|---|---|---|---|
| Borophene | **a** | 20.26 | 0.08 | Phonon instability |
|  | **b** | 12.98 | 0.15 | Elastic instability |
|  | Biaxial | 14.75 | 0.08 | Phonon instability |
| Graphene[12] | Zigzag | 40.41 | 0.266 | Elastic instability |
|  | Armchair | 36.74 | 0.194 | Elastic instability |
| Silicene[32] | Armchair | 7.59 | 0.17 | Elastic instability |
|  | Zigzag | 5.26 | 0.136 | Elastic instability |
|  | Biaxial | 6.76 | 0.16 | Elastic instability |
| BP[25] | Zigzag | 9.99 | 0.27 | / |
|  | Armchair | 4.44 | 0.33 | / |
| $MoS_2$[22] | Armchair | 14.75 | 0.256 | Phonon instability |
|  | Zigzag | 9.59 | 0.18 | Elastic instability |
|  | Biaxial | 14.63 | 0.195 | Phonon instability |

In summary, we investigated the ideal tensile strength and critical strain of monolayer borophene through first principles calculations. We found that the ideal tensile strength of borophene demonstrates significant anisotropy. Monolayer borophene can withstand stress up to 20.26 N/m and 12.98 N/m in **a** and **b** directions, respectively. The corresponding critical strains are 8% (**a** direction) and 15% (**b** direction). Compared to other 2D materials, we found borophene is a hard and brittle 2D material. It was found that the bucking height of borophene increases with strain applied in **b** direction. This means 2D borophene has an out-of-plane negative Poisson's ratio, which effectively holds the strong $\sigma$ bonds lying along **a** direction and makes the boron sheet show superior mechanical flexibility along **b** direction. Furthermore, the phonon instability of monolayer borophene was studied and the results mean that under uniaxial tension along **b** direction, it is attributed to elastic instability, while along **a** direction and biaxial tension the failure mechanism is phonon instability and such an instability is dictated by out-of-plane acoustical mode.

**Acknowledgements**

This work was supported by the National Key Project for Basic Research of China (Grant nos 2015CB921600, 2012CB921900), Natural Science Foundation of China (Grant nos 11547030, 11525417, 11374142 and 11174152), the Opening Project of Shanghai Key Laboratory of High Temperature Superconductors (14DZ2260700), and the Program for New Century Excellent Talents in University (Grant No. NCET-12-0278).